\documentclass[aps,prd,preprint,superscriptaddress,showpacs]{revtex4}
\begin{document}

\title{One-loop approximation of M{\o}ller scattering in\\ Krein-space quantization}

\author{F. Payandeh}
\email{payandeh@aut.ac.ir} \affiliation{Physics Department,
Amirkabir University of Technology, Tehran 15914,Iran}
\affiliation{Physics Department, Payame Noor University, Tehran
19395-4697, Iran}
\author{M. Mehrafarin}
\email{mehrafar@aut.ac.ir} \affiliation{Physics Department,
Amirkabir University of Technology, Tehran 15914,Iran}
\author{M. V. Takook}
\email{takook@razi.ac.ir}
\affiliation{Physics Department, Razi University, Kermanshah,
Iran}

\date{\today}

\vspace{15pt}

\begin{abstract}
It has been shown that the negative-norm states necessarily appear
in a covariant quantization of the free minimally coupled scalar
field in de Sitter spacetime \cite{De57,Tak17}. In this processes
ultraviolet and infrared divergences have been automatically
eliminated \cite{Tak16}. A natural renormalization of the one-loop
interacting quantum field in Minkowski spacetime ($\lambda\phi^4$)
has been achieved through the consideration of the negative-norm
states defined in Krein space. It has been shown that the
combination of quantum field theory in Krein space together with
consideration of quantum metric fluctuation, results in quantum
field theory without any divergences \cite{Rouh}. Pursuing this
approach, we express Wick's theorem and calculate M{\o}ller
scattering in the one-loop approximation in Krein space. The
mathematical consequence of this method is the disappearance of
the ultraviolet divergence in the one-loop approximation.
\end{abstract}


\maketitle

\section{Introduction}
In view of the appearance of infrared divergence in the two-point
function for the minimally coupled scalar field in de Sitter
spacetime, a new method of field quantization called ``Krein QFT"
has been proposed, which uses negative-norm states
\cite{De57,Tak17,Tak16,Rouh}.

Consideration of the negative-norm states was proposed by Dirac in
1942 \cite{D180}. In 1950 Gupta applied the idea in QED
\cite{Gupt}. The presence of higher derivatives in the Lagrangian
also leads to ghosts; states with negative norm \cite{Haw}. The
auxiliary negative-norm states were primarily introduced in de
Sitter spacetime to achieve covariant quantization. However, their
presence has other consequences, too. For example, in QED the
negative-energy photon disappears \cite{Gupt}, and in de Sitter
spacetime the infrared divergence of minimally coupled scalar
field is eliminated \cite{ta1}. In similarity with linear gravity,
this divergence does not manifest itself in the quadratic part of
the effective action in the one-loop approximation. This means
that the pathological behavior of the graviton propagator is gauge
dependent and so should not appear in an effective way as a
physical quantity \cite{A462,G44}. Ignoring the positivity
condition (for norm and energy), similar to Gupta-Bleuler
quantization of the electrodynamics in Minkowski spacetime, the
quantization of free boson and spinor fields has been performed in
Krein space \cite{PaTa,Spinor}. Following this scheme, the normal
ordering procedure is rendered useless since the vacuum energy
remains convergent \cite{PaTa,Spinor}.

It is worthwhile to note that the unphysical (negative-energy)
states for boson fields have both negative and positive norms and
are, therefore, defined in Krein space. In Gupta-Bleuler
quantization, the unphysical states with positive and negative
norms are introduced in order to preserve the Lorentz invariance.
The negative-norm states appear due to the sign of the Minkowskian
metric. However, in Krein space quantization the appearance of the
additional negative-norm states owes itself to the
negative-frequency solutions. The two sets of solutions for boson
fields in Krein space are complex conjugates. However, for spinor
fields the unphysical states are positive-norm states (moving
forward in time). The unphysical positive-norm states of spinor
fields are different from physical antiparticles. The latter,
although having negative energy, move backward in time whereas the
former move forward in time and are not observable. In other
words, in the spinor case the unphysical state may be regarded as
the unphysical particle and antiparticle in the inverse time
direction. It is noteworthy that the unphysical states play no
role in the physical world and they are just used as a
mathematical tool.

The most interesting result of the new construction is the
convergence of the Green's function at large distances, which
means that the infrared divergence is gauge dependent
\cite{Tak17,Tak16}. The ultraviolet divergence in the stress
tensor disappears as well, so that the quantum free fields are
automatically renormalized by this method. The role of unphysical
states appears in the method as a natural renormalization tool in
the one-loop approximation. It is worthwhile to note that via this
method, a natural renormalization of the following problems are
attained:

\noindent $\bullet$ The massive free field in de Sitter spacetime
\cite{Tak17}.

\noindent $\bullet$ The graviton two-point function in de Sitter
spacetime \cite{Takwig}.

\noindent $\bullet$ The one-loop effective action for scalar field
in a general curved spacetime \cite{Tak14}.

\noindent $\bullet$ Tree level scattering amplitude for scalar
field with one graviton exchange in de Sitter spacetime
\cite{Tak68}.

\noindent $\bullet$ Casimir effect in Krein space quantization
\cite{Casimir}.

\noindent $\bullet$  Free fields Quantization in Krein space
\cite{PaTa,Spinor}.

Following the above works, in this paper we express Wick's theorem
and calculate the M{\o}ller (electron-electron) scattering matrix
$ S$ in the one-loop approximation in Krein space. Again, it is
seen that the presence of negative-norm states plays the role of
an automatic renormalization tool for the theory of quantum
fields. A number of works already published about Krein space and
quantum field theories with indefinite metric could be found in
\cite{GHR,Hofman1,Hofman2}.

\section{Krein QFT Calculation}
The origin of divergence in QFT lies in the singular character of
the Green's function at short relative distances. In Krein QFT,
although the Green's function is changed due to the presence of
negative-frequency states, we see that these unphysical states
disappear as far as observable average values are concerned
\cite{PaTa,Tak11}.

One of the interesting results of constructing field theory in
Krein space is that all ultraviolet divergences of QFT with the
exception of the light cone singularity are removed. It was
conjectured long ago \cite{D29,D13} that quantum metric
fluctuations does smear out the light cone singularities, but it
does not remove other ultraviolet divergences. Indeed, it has been
shown that quantum metric fluctuations remove the singularities of
Green's functions on the light cone \cite{Ford}.

In a previous work \cite{Rouh}, it has been established that the
combination of QFT in Krein space together with consideration of
quantum metric fluctuations results in a QFT without any divergence
($T$ means
time-ordered):
\begin{equation}
<G_T(x-x^\prime)>=-\frac{1}{8\pi}\sqrt{\frac{\pi}{2<\sigma_1^2>}}\exp(-\frac{\sigma_0^2}{2<\sigma_1^2>})
+\frac{m^2}{8\pi}\
\theta(\sigma_0)\frac{J_1(\sqrt{2m^2\sigma_0})}{\sqrt{2m^2\sigma_0}}
\label{1}
\end{equation}
where $
2\sigma_0=\eta_{\mu\nu}(x^\mu-x^{\prime\mu})(x^\nu-x^{\prime\nu})$,
and $<\sigma_1^2>$ is related to the density of gravitons. When
$\sigma_0=0$, due to the metric quantum fluctuation
$<\sigma_1^2>\neq0$, we have
$$
<G_T(0)>=-\frac{1}{8\pi}\sqrt{\frac{\pi}{2<\sigma_1^2>}}+\frac{m^2}{8\pi}.
$$

The field operators and their associated divergence-free Green's
functions are \cite{PaTa} ($K$  shall, hereafter, always stand for quantities in Krein space),

\noindent (i) Klein-Gordon field:
$$
 \phi_K(x)=\frac{1}{\sqrt 2}\int d^3{\vec k}\ [( a_{\vec
k}+ b_{\vec k}^\dag) u_p(k,x) + ( a_{\vec k}^\dag+
b_{\vec k}) u_n(k,x)]
$$
where
$$
u_p(k,x)=\frac{e^{-ik.x}}{\sqrt{(2\pi)^3 2\omega_{\vec k}}},
\ \ u_n(k,x)=\frac{e^{ik.x}}{\sqrt{(2\pi)^3 2\omega_{\vec k}}}, \ \ \omega_{\vec k}=\sqrt{{\vec k}.{\vec k}+m^2}
$$
and the Green's function is given by (\ref {1}).

\noindent (ii) Maxwell field:
$$
A^K_\mu (x)=\frac{1}{\sqrt 2}\int d^3{\vec k}  \sum
_{\lambda=0}^{3} \epsilon^\lambda_\mu({\vec k}) [ (  a^\lambda_{\vec k}
+ b^{\lambda\dag}_{\vec k}) \ u_p(k,x) + (
a_{\vec k}^{\lambda\dag}+ b_{\vec k}^\lambda)u_n(k,x)]
$$
and
$$
D^T _ {\mu\nu}(x,x^\prime)=-\eta_{\mu\nu}G_T(x,x^\prime)
$$
\noindent is the time-ordered propagator.

\noindent (iii) Dirac field ( Krein space):
$$
\psi_K(x)=\frac{1}{\sqrt 2}\int d^3{\vec k} \sum_{s=1,2} [(
b_{{\vec k}s}+ c_{{\vec k}s}^\dag){\cal U}^s(k,x)+( d_{{\vec
k}s}^\dag+ a_{{\vec k}s}){\cal V}^s(k,x)]
$$

\noindent where

$$
{\cal U}^s(k,x)=\sqrt{\frac{m}{(2\pi)^3\omega_{\vec k}}} \,\
u^s(\vec k) e^{-ik.x} \,\ \,\ \,\  (\mbox{positive energy})
$$

$$
{\cal V}^s(k,x)=\sqrt{\frac{m}{(2\pi)^3\omega_{\vec k}}} \,\
v^s(\vec k) e^{ik.x} \,\ \,\ \,\  (\mbox{negative energy})
$$

\noindent and

$$
S_T(x,x^\prime)=(i\not\partial+m) G_T(x,x^\prime)
$$
is the time-ordered propagator, that is
\begin{eqnarray}
S_T(x,x^\prime)=\frac{1}{8\pi}\ i\gamma^\mu (x_\mu-x_\mu^\prime)\{\sqrt{\frac{\pi}{2<\sigma_1^2>}}\  e^{-\frac{\sigma_0^2}{2<\sigma_1^2>}}[ \frac{\sigma_0}{<\sigma_1^2>}  +m^2\frac{J_1(\sqrt{2m^2\sigma_0})}{\sqrt{2m^2\sigma_0}}]\nonumber \\
+\frac{m}{2\sqrt{2}}\ \theta(\sigma_0) [\sqrt{2m^2\sigma_0} \
J_0(\sqrt{2m^2\sigma_0})-2J_1(\sqrt{2m^2\sigma_0})] \} \nonumber \\
+\frac{m}{8\pi}\
[-\sqrt{\frac{\pi}{2<\sigma_1^2>}}e^{-\frac{\sigma_0^2}{2<\sigma_1^2>}}+m^2
\theta(\sigma_0)\frac{J_1(\sqrt{2m^2\sigma_0})}{\sqrt{2m^2\sigma_0}}]\nonumber
\end{eqnarray}

\section{Wick's Theorem in Krein Space}
Let $A$ and $B$ be two linear operators in Krein space:
\begin{eqnarray}
A_K(x)=A_p(x)+A_n(x)=(A_p^{+} + A_p^{-})+(A_n^{+} + A_n^{-}) \nonumber \\
B_K(x)=B_p(x)+B_n(x)=(B_p^{+} + B_p^{-})+(B_n^{+} + B_n^{-})
\nonumber
\end{eqnarray}
where $A_p(x)$, $B_p(x)$ are the physical (positive-frequency) and
$A_n(x)$, $B_n(x)$ the unphysical (negative-frequency) parts of
the operators. We have
\begin{eqnarray}
A_p^{+} |0>=0  ,\ \ A_n^{+} |0>=0  ,\ \
A_p^{-} |0>=|1_{\vec k}>  ,\ \ A_n^{-} |0>=|\bar1_{\vec k}>\nonumber \\
B_p^{+} |0>=0   ,\ \ B_n^{+} |0>=0  ,\ \
B_p^{-} |0>=|1_{\vec k}>   ,\ \ B_n^{-} |0>=|\bar1_{\vec k}>\nonumber \\
A_K B_K = :A_K B_K: + [A_p^{+},B_p^{-}] + [A_n^{+},B_n^{-}] \label{2}\\
<0|A_K B_K|0>= [ A_p^{+},B_p^{-}]+ [A_n^{+},B_n^{-}] \label{3}
\end{eqnarray}
Comparing (\ref{2}) and (\ref{3}) yields
$$
A_K B_K = :A_K B_K: + <0|A_K B_K|0>
$$
and
\begin{equation}
<0|T[A_K(x_1)B_K(x_2)]|0>= \widehat{ [A_K(x_1)B_K(x_2)]} \label{4}
\end{equation}
Since the contractions of physical and unphysical parts are zero,
(\ref{4}) reads
$$
<0|T[A_K(x_1)B_K(x_2)]|0>=\widehat{[A_p(x_1)B_p(x_2)]} +
\widehat{[A_n(x_1)B_n(x_2)]}
$$
Thus, the contraction of two operators in Krein space splits into
the contraction of physical parts, and the contraction of
unphysical parts. We have
\begin{eqnarray}
<0|T[\phi_K(x_1)\phi_K(x_2)]|0>= iG_T(x_1,x_2)\ \ \ \ \ \ \ \ \ \ \ \ \ \ \ \ \ \nonumber \\
<0|T[ A^K_\mu (x_1) A^K_\nu(x_2) ]|0>
=iD^T_{\mu\nu}(x_1,x_2)=-i\eta_{\mu\nu}G_T(x_1,x_2) \nonumber \\
<0|T[ \psi_K(x_1) \psi_K(x_2)]|0> =iS_T(x_1,x_2)=i(i\not\partial+m)
G_T(x_1,x_2) \nonumber
\end{eqnarray}
and Wick's theorem for the time-ordered product of some linear
operators  $ A_K$, $ B_K$, ..., $ Z_K$ in Krein space is expressed
by
\begin{eqnarray}
T[ A_K(x_1)B_K(x_2)...]=[:A_p B_p...Z_p: + :A_n B_n...Z_n:]
+ [\widehat{ (A_p B_p)} :C_p...Z_p : \nonumber \\
+\widehat{ (A_n B_n)} :C_n...Z_n:]+...
+ [\widehat{ (A_p B_p)} \widehat{ (C_p D_p)}:E_p...Z_p: \nonumber \\
+\widehat{(A_n B_n)} \widehat{ (C_n D_n)}:E_n...Z_n:] +... \nonumber
\end{eqnarray}

\section{Feynman Rules in Krein space}
In Krein space, the graphical ``translation rules" introduced by
Feynman  can be written as \cite{Greiner}:

\noindent 1. Each point of interaction $x$ is associated with a
{\it vertex}. This corresponds to the algebraic factor
$-i\gamma^\mu$.

\noindent 2. Each electron field operator is associated with an
external fermion line and an external unphysical fermion line
glued to a vertex. The physical lines are full while the
unphysical lines are dashed. These lines carry an arrow; however,
if the physical lines are forward in time, the unphysical lines
run backward in time. The field operators in Krein space are
\begin{eqnarray}
\psi_K(x)=\psi_p(x)+\psi_n(x)=[\psi_p^{+}(x)+\psi_p^{-}(x)]+[\psi_n^{+}(x)+\psi_n^{-}(x)]\nonumber \\
\bar\psi_K(x)=\bar\psi_p(x)+\bar\psi_n(x)=[\bar\psi_p^{+}(x)+\bar\psi_p^{-}(x)]+[\bar\psi_n^{+}(x)+\bar\psi_n^{-}(x)]\nonumber \\
A^K_\mu(x)=A^p_\mu(x)+A^n_\mu(x)=[A_{\mu_p}^{+}(x)+A_{\mu_p}^{-}(x)]+[A_{\mu_n}^{+}(x)+A_{\mu_n}^{-}(x)]\nonumber
\end{eqnarray}
The following associations are made:

\noindent (i)$\ \psi_p^{+}(x)$: a line pointing upward ending at $x$
(electron absorption)

\noindent (ii)$\ \psi_n^{+}(x)$: a dashed line pointing downward
starting at $x$ (unphysical positron absorption)

\noindent (iii)$\ \psi_p^{-}(x)$: a line pointing downward ending at
$x$ (positron emission)

\noindent (iv)$\ \psi_n^{-}(x)$: a dashed line pointing upward
starting at $x$ (unphysical electron emission)

\noindent (v)$\ \bar\psi_p^{+}(x)$: a line pointing downward
starting at $x$ (positron absorption)

\noindent (vi)$\ \bar\psi_n^{+}(x)$: a dashed line pointing upward
ending at $x$ (unphysical electron absorption)

\noindent (vii)$\ \bar\psi_p^{-}(x)$: a line pointing upward
starting at $x$ (electron emission)

\noindent (viii)$\ \bar\psi_n^{-}(x)$: a dashed line pointing
downward ending at $x$ (unphysical positron emission)

\noindent 3. Each photon field operator is associated with a
wiggly external photon line and each unphysical photon line is
associated with a dashed wiggly external photon line:

\noindent (i)$\ A_{\mu_p}^{+}(x)$: a line starting at $x$ pointing
downward (photon absorption)

\noindent (ii)$\ A_{\mu_n}^{+}(x)$: a dashed line ending at $x$
pointing upward (unphysical photon absorption)

\noindent (iii)$\ A_{\mu_p}^{-}(x)$: a line starting at $x$ pointing
upward (photon emission)

\noindent (iv)$\ A_{\mu_n}^{-}(x)$: a dashed line ending at $x$
pointing downward (unphysical photon emission)

\noindent 4. The contraction of two fermion operators in Krein
space,
$$
\widehat{[\psi_K(x_1)\psi_K(x_2)]}=iS_T(x,x^\prime),
$$
is associated with a directed internal fermion line from $x_1$ to
$x_2$, and a directed unphysical fermion line from $x_2$ to $x_1$.

\noindent 5. The contraction of two photon operators in Krein
space,
$$
\widehat{ [A^K_\mu(x_1) A^K_\nu(x_2)]}=iD^T _
{\mu\nu}(x_1,x_2),
$$
is associated with a wiggly internal photon line connecting $x_1,
x_2$ and a dashed wiggly internal unphysical photon line
connecting $x_1, x_2$. It is remarkable that the photon lines have
no sense of direction (they bear no arrow) since the photon ``is
its own antiparticle". This is reflected in the symmetry of the
photon propagator.

Since the physical and unphysical states are orthogonal to each
other, the expectation values of unphysical operators on the
physical initial and final states will be zero. Therefore, to
evaluate an $S$ matrix element of $n$th order, all topologically
distinct graphs with $n$ vertices and within the desired
configuration of external physical lines are drawn. Each vertex is
assigned a coordinate variable $x_i$, and the Feynman rules of QED
in coordinate space in Krein-space quantization read:

\noindent 1. Vertex: $-ie{(\gamma_\mu)_{\alpha\beta}}$

\noindent 2. Internal photon and unphysical photon lines: $iD^T _
{\mu\nu}(x_k,x_l)$

\noindent 3. Internal fermion and unphysical fermion lines: $
iS_T(x_k,x_l)$

\noindent 4. External fermion line:

$u^s(\vec k) u_p(k,x)$ (incoming electron)

$\bar v^s(\vec k) u_p(k,x)$ (incoming positron)

$\bar u^s(\vec k) u_n(k,x)$ (outgoing electron)

$v^s(\vec k) u_n(k,x)$ (outgoing positron)

\noindent 5. External photon line:

$\epsilon^\lambda_\mu(\vec k) u_p(k,x)$ (incoming photon)

$\epsilon^\lambda_\mu(\vec k) u_n(k,x)$ (outgoing photon)

\noindent 6. All coordinates $x_i$ are integrated over. The
integration can be carried out owing to the simple plane-wave
factors.

\noindent 7. Each closed fermion loop leads to a factor $-1$.

It is seen that in calculations the unphysical states are eliminated
in the external legs and are introduced only in the propagators.
They eliminate the divergences of the theory automatically and
without any renormalization procedure. The physical meaning of the
negative-frequency states is not clear. However, by analogy with the
standard QFT, one can interpret them as negative-energy unphysical
particles (or antiparticles) by changing the time direction.

\section{M{\o}ller Scattering in Krein Space}
In M{\o}ller scattering, two electrons with momenta and spins
$k_1,s_1$ and $k_2,s_2$ in the initial state are scattered into
the final state with $k_1^\prime,s_1^\prime$ and
$k_2^\prime,s_2^\prime$. According to the Wick expansion of the $
S$ operator, for the one-loop approximation of M{\o}ller
scattering in Krein-space quantization, we have up to fourth order
\begin{eqnarray}
 S=\frac{(-ie)^4}{4!} \int d^4x_1\ d^4x_2\ d^4x_3\ d^4x_4 :\bar
\psi_K^{-}(x_1) \gamma^\mu  \bar\psi_K^{+}(x_1)\
iD^T_{\mu\alpha}(x_1-x_3)\ (-Tr)[\nonumber \\
iS_T(x_4-x_3)\gamma^\alpha iS_T(x_3-x_4)\gamma^\beta]\
iD^T_{\beta\nu}(x_2-x_4)\ \bar \psi_K^{-}(x_2) \gamma^\nu
\bar\psi_K^{+}(x_2):\nonumber
\end{eqnarray}
The $S$-matrix element of M{\o}ller scattering is given by
$$
S_{fi}=<0|b_{{\vec k}_2^\prime s_2^\prime} b_{{\vec k}_1^\prime s_1^\prime}\
 S\  b_{{\vec k}_1s_1}^\dag b_{{\vec k_2}s_2}^\dag |0>
$$
Insert the field operators of Krein space and discard the products
of the physical and unphysical states because of the separation of
their spaces; $S_{fi}$ splits into the sum of two physical and
unphysical terms. Because of the orthogonality of the physical and
unphysical states, the expectation values of unphysical operators
on physical $|i>$ and $<f|$ states are zero. Therefore the
unphysical term of $S_{fi}$ vanishes, and the final result for the
$S$-matrix element of M{\o}ller scattering in the one-loop
approximation in Krein space reads:

\begin{eqnarray}
S_{fi}=\frac{(-ie)^4}{2!}\frac{1}{(2\pi)^6} \int d^4x_1\ d^4x_2\
d^4x_3\ d^4x_4\ \sqrt{\frac{m}{ \omega_{{\vec k}_1}}}
\sqrt{\frac{m}{ \omega_{{\vec k}_2}}} \sqrt{\frac{m}{
\omega_{{\vec k}_1^\prime}}} \sqrt{\frac{m}{ \omega_{{\vec
k}_2^\prime}}}\
iD^T_{\mu\alpha}(x_1-x_3)(-Tr)[ \nonumber \\
iS_T(x_4-x_3) \gamma^\alpha iS_T(x_3-x_4) \gamma^\beta]
iD^T_{\beta\nu}(x_2-x_4) [e^{i(k_2^\prime-k_2).x_2}
e^{i(k_1^\prime-k_1).x_1}\nonumber \\
\bar u^{ s_2^\prime}(\vec k_2^\prime) \gamma^\mu u^{ s_2}(\vec k_2) \bar
u^{ s_1^\prime}(\vec k_1^\prime) \gamma^\nu u^{s_1}(\vec k_1)-
e^{i(k_2^\prime-k_1).x_2}
e^{i(k_1^\prime-k_2).x_1}\nonumber \\
\bar u^ {s_2^\prime }(\vec k_2^\prime) \gamma^\mu u^{s_1} (\vec k_1) \bar
u^{s_1^\prime}(\vec k_1^\prime) \gamma^\nu  u^{s_2}(\vec
k_2)]\nonumber
\end{eqnarray}
which is free of any divergence. Of course, it is not enough to
say that some amplitudes are finite. It is also necessary to show
that it has something to do with the real stuff and processes
which occur in nature. So, We are going to calculate it in the
forthcoming papers.

\section{Conclusion}
The negative-frequency solutions of the field equation are
indispensable for the covariant quantization of minimally coupled
scalar field in de Sitter spacetime. Contrary to the Minkowski
spacetime, discarding the negative-frequency states in this case
breaks the de Sitter invariance. In other words, for restoration
of de Sitter invariance, one must take the negative-frequency
states into account in the Krein-space quantization. Free boson
and spinor fields are investigated in Krein space quantization. In
the present paper the method has been applied to QED, and the
one-loop approximation of M{\o}ller scattering is investigated in
Krein space. Once again it is found that the theory is
automatically renormalized.

\end{document}